# Comment on [arXiv:2306.07869v1](arXiv:2306.07869v1) Light-induced melting of competing stripe orders without introducing superconductivity in $La_{1.875}Ba_{0.125}CuO_4$

D. Nicoletti[1]*, M. Buzzi[1], M. Först[1], and A. Cavalleri[1,†]

[1]*Max Planck Institute for the Structure and Dynamics of Matter, Hamburg, Germany*
*daniele.nicoletti@mpsd.mpg.de*
[†]*andrea.cavalleri@mpsd.mpg.de*

In the manuscript [arXiv:2306.07869v1](arXiv:2306.07869v1), N. L. Wang and co-authors report the results of a near-infrared pump / terahertz probe study in the stripe-ordered cuprate $La_{1.875}Ba_{0.125}CuO_4$. They measured a change in optical conductivity, but did not find signatures of transient superconductivity. From this observation they extrapolate that in all cuprates in which striped states have been excited with light, there must be no light-induced superconductivity. They conclude that "transient superconductivity cannot be induced by melting of the competing stripe orders with pump pulses whose photon energy is much higher than the superconducting gap of cuprates." Here we show that this extrapolation is unwarranted. First, the absence of light-induced superconductivity in this particular compound was already reported in a previous paper, which instead showed positive evidence for $La_{1.885}Ba_{0.115}CuO_4$. In addition, the experiment discussed here used photo-excitation with too low fluence and at a suboptimal wavelength. More broadly, a negative result in one compound is rarely compelling indication of the absence of an effect in an entire class of materials.

In Reference 1, N. L. Wang and co-authors report a study of photo-induced dynamics in the stripe-ordered cuprate $La_{1.875}Ba_{0.125}CuO_4$ (LBCO 12.5%) after excitation in the near infrared. This work connects to a series of pump-probe experiments [2-4] carried out in recent years by different groups, including our own, on a number of similar "striped" cuprates. In these, it was shown how photoexcitation in various spectral ranges could result in a transient perturbation of the stripe phase and, consequently, in the appearance of a coherent, Josephson-like terahertz response along the *c*-axis, for temperatures far above the equilibrium superconducting $T_C$.

Specifically, the first such result was reported in 2011 on $La_{1.675}Eu_{0.2}Sr_{0.125}CuO_4$ (LESCO) upon excitation in the mid infrared, resonant with an in-plane Cu-O stretching mode [2]. This was then followed by analogous experiments on $La_{1.885}Ba_{0.115}CuO_4$ (LBCO 11.5%), in this case with near-infrared charge excitation at 800-nm wavelength, first reported by our group [3] and then reproduced by R. D. Averitt and collaborators in Ref. 4.

These previous results share two features. First, a sharp mode appeared in the transient *c*-axis reflectivity and loss function from a featureless background at equilibrium. Second, the imaginary part of the optical conductivity increased, became positive and exhibited a $1/\omega$-like divergence at low frequencies. Both of these properties were found to be virtually identical to the superconducting response at equilibrium and were therefore interpreted as signatures of optically-induced superconductivity.

In [arXiv:2306.07869v1](arXiv:2306.07869v1) [1], the authors do not observe these two features in the related compound LBCO 12.5% excited at a different wavelength of 1.3 μm. Figure 1 reveals the qualitative difference between the *c*-axis reflectivity and imaginary conductivity reported for LBCO 11.5% by our group [3] (panels a,b) and those of LBCO 12.5% measured by N. L. Wang *et al.* [1] (panels c,d). In LBCO 11.5% a sharp reflectivity edge develops from a featureless background (Fig. 1a) and a divergence appears in $\sigma_2(\omega)$ at low frequencies (Fig. 1b), which is a signature of "dissipation-less" transport. LBCO 12.5% instead displays a reshaping of the reflectivity over a broad frequency range and no $\sigma_2(\omega)$ divergence, a behaviour that is well reproduced by a Drude model with finite scattering.

Note that the absence of a coherent superconducting-like response in photo-excited LBCO 12.5% had already been reported by our group in 2014 [3], and discussed in terms of a reduced photo-susceptibility at this doping level with respect to 11.5%. At this ~1/8 hole concentration, in fact, the stripe order is almost "static" and exhibits its maximum correlation length [5].

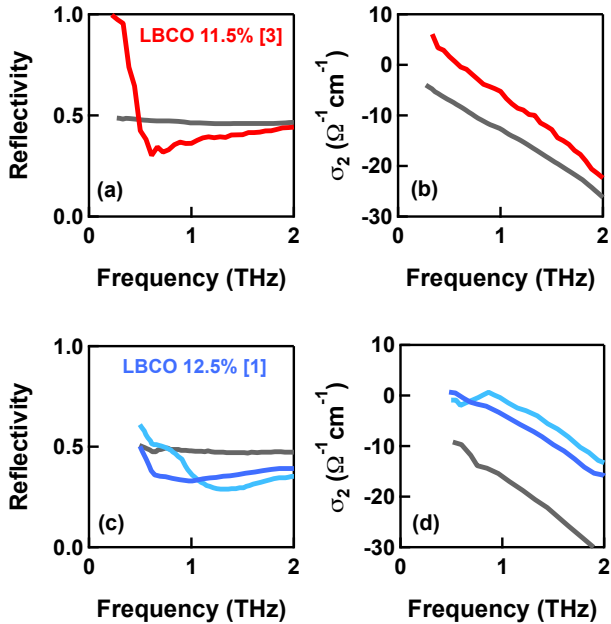

Figure 1: (a) *c*-axis reflectivity and (b) imaginary part of the optical conductivity measured in LBCO 11.5% before (grey) and after (red) photoexcitation with *c*-axis polarized 800-nm pulses at 2 mJ/cm², at the peak of the light-induced response. Data reproduced from Ref. 3. (c,d) Same quantities as in (a,b) measured in LBCO 12.5% upon excitation with 1.3 μm pulses at 1 mJ/cm² fluence, reproduced from Ref. 1. Here the transient spectra are shown at 0.9 ps (dark blue) and 2.1 ps (light blue) pump-probe delay.

Another important issue is the choice of excitation parameters in the experiment by N. L. Wang *et al*. [1]. Figure 2 shows the wavelength and pump fluence dependence for the photoinduced superconducting-like response in LBCO 11.5% previously reported by our group in Ref. 6. Therein, we note that for optimal near-infrared charge excitation the pump wavelength must be tuned near 800 nm, being close to the charge transfer resonance peak in the absorption spectrum. For longer wavelengths (see data at 2 μm pump), the optically-driven superconductivity process is far less efficient.

It is evident here that, even if the authors of Ref. 1 had measured on a more suitable compound, with their suboptimal choice of pump wavelength (1.3 μm) the applied fluence of 1 mJ/cm² would have been insufficient to observe a fully saturated superconducting-like response.

Also, to the best of our knowledge, a charge-order melting in LBCO 12.5% under these experimental conditions has never been demonstrated. The only experiments we are aware of have measured a stripe melting dynamics by soft x-ray scattering in the same compound upon excitation along the Cu-O planes in the mid infrared [7] and at 800-nm wavelength [8]. The choice of title of the paper is therefore arbitrary.

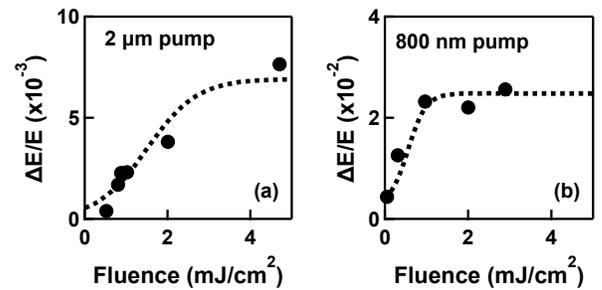

Figure 2: Differential time-domain transients of the reflected THz electric-field peak, measured in LBCO 11.5% 1.5 ps after photoexcitation and plotted as a function of pump fluence for different excitation wavelengths. The black lines are sigmoid function fits, returning saturation fluences of 3.0 and 1.1 mJ/cm² for (a) 2-μm and (b) 800-nm pump, respectively. Data reproduced from Ref. 6.

In conclusion, if the authors' purpose were to critically reconsider previous experiments of optically-driven superconductivity in striped cuprates, we believe that their choice of both material and photo-excitation conditions is far from optimal. We suggest that these aspects should have been mentioned, alongside proper referencing of Refs. 3,6.